\documentclass[twocolumn]{aastex631}  

\usepackage{amsmath}
\usepackage{graphicx}
\usepackage{txfonts}

\usepackage{natbib}
\begin{document}

\title{Glossy Silicate Clouds on the Scorched Dayside of LTT9779b}

\author[0000-0001-8018-0264]{Suman Saha}
\affiliation{Instituto de Estudios Astrofísicos, Facultad de Ingeniería Ciencias, Universidad Diego Portales, Av. Ejército Libertador 441, Santiago, Chile}
\affiliation{Centro de Excelencia en Astrofísica y Tecnologías Afines (CATA), Camino El Observatorio 1515, Las Condes, Santiago, Chile}
	
\correspondingauthor{Suman Saha}
\email{suman.saha@mail.udp.cl}

\author[0000-0003-2733-8725]{James S. Jenkins}
\affiliation{Instituto de Estudios Astrofísicos, Facultad de Ingeniería Ciencias, Universidad Diego Portales, Av. Ejército Libertador 441, Santiago, Chile}
\affiliation{Centro de Excelencia en Astrofísica y Tecnologías Afines (CATA), Camino El Observatorio 1515, Las Condes, Santiago, Chile}

\author[0000-0002-2072-6541]{Jonathan Brande}
\affiliation{Department of Astronomy, University of Maryland, PSC 1113, 4296 Stadium Dr., College Park, MD 20742, USA}

\author[0000-0002-0786-7307]{Reza Ashtari}
\affiliation{Johns Hopkins APL, 11100 Johns Hopkins Rd, Laurel, MD 20723, USA}

\author{Ian J. M. Crossfield}
\affiliation{Department of Physics and Astronomy, University of Kansas, 1082 Malott Hall, 1251 Wescoe Hall Dr, Lawrence, KS 66045, USA}

\author[0000-0003-1131-8922]{Sarah Stamer}
\affiliation{Department of Physics and Astronomy, University of New Mexico, 210 Yale Blvd NE, Albuquerque, NM 87106, USA}

\author[0000-0003-2313-467X]{Diana Dragomir}
\affiliation{Department of Physics and Astronomy, University of New Mexico, 210 Yale Blvd NE, Albuquerque, NM 87106, USA}

\author[0000-0002-7352-7941]{Kevin B. Stevenson}
\affiliation{Johns Hopkins APL, 11100 Johns Hopkins Rd, Laurel, MD 20723, USA}

\author[0000-0001-9521-6258]{Vivien Parmentier} 
\affiliation{Universite Cote d'Azur, Av. Valrose, 06000 Nice, France}

\author[0000-0001-5442-1300]{Thomas M. Evans-Soma}
\affiliation{School of Science, University of Newcastle, Callaghan, NSW, Australia}

\author[0000-0002-6939-9211]{Tansu Daylan}
\affiliation{Department of Physics and McDonnell Center for the Space Sciences, Washington University, St. Louis, MO 63130, USA}

\author[0000-0002-6980-052X]{Hayley Beltz}
\affiliation{Department of Physics and Astronomy, University of Kansas, 1082 Malott Hall, 1251 Wescoe Hall Dr, Lawrence, KS 66045, USA}

\author[0000-0002-2341-3233]{Emma Esparza-Borges}
\affiliation{Instituto de Astrof\'isica de Canarias, C. V\'ia
L\'actea, San Crist\'obal de La Laguna 38205, Spain}

\accepted{for publication in The Astrophysical Journal Letters}

\begin{abstract}
Discovered deep within the “Neptunian desert", LTT9779b remains the only known ultra-hot Neptune, prompting significant speculation regarding its unique formation and evolutionary history. Its exceptionally high geometric albedo has previously been attributed either to the presence of clouds or to an extremely metal-rich atmosphere. Here, we present a comprehensive panchromatic analysis of its dayside atmosphere using JWST NIRISS and NIRSpec/G395H observations to characterize its atmospheric structure and composition. Leveraging the exceptional signal-to-noise ratio (S/N) in the observed spectra, we report a 3-to-5$\sigma$ detection of dayside clouds, with strong evidence for Mg$_2$SiO$_4$(s) (silicate) condensation. This constitutes the first statistically significant detection of clouds on the dayside of a Neptunian-mass exoplanet. We demonstrate that a highly reflective cloud deck, rather than an extremely high-metallicity atmosphere, is the most likely explanation for the planet’s anomalously high optical albedo. Furthermore, our atmospheric retrievals yield robust detections of both CO ($\sim$4.88$\sigma$) and CO$_2$ ($\sim$8.76$\sigma$), while providing tentative constraints on the H$_2$O abundance and upper limits on SiO, TiO, and VO. Finally, our analysis places a robust constraint on the C/O ratio of 0.984 $\pm$ 0.019. This aligns LTT9779b with other known ultra-hot Jupiters exhibiting super-solar C/O ratios, suggesting a broader trend driven by the sequestration of oxygen-bearing condensates in ultra-hot atmospheres.
\end{abstract}

\section{Introduction}\label{sec:sec1}

\begin{figure}
\centering
\includegraphics[width=0.48\textwidth]{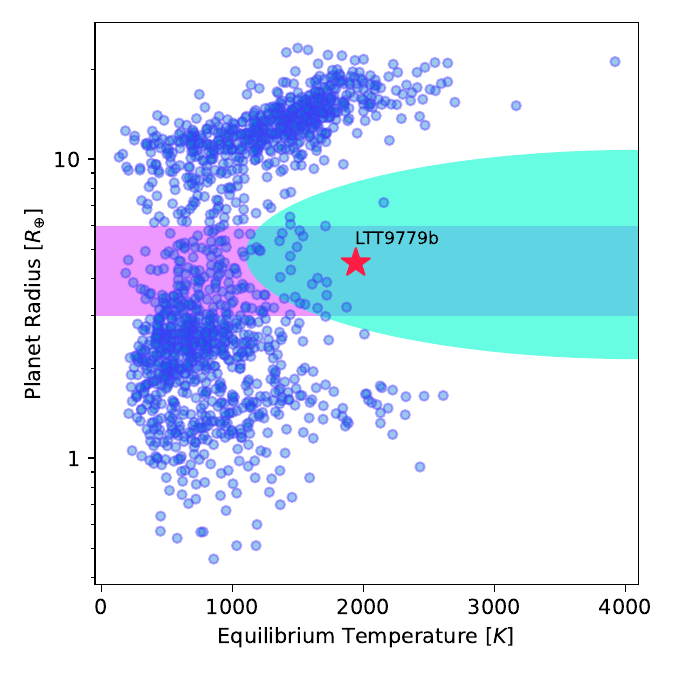}
\caption{Equilibrium temperature vs. planetary radius for known transiting exoplanets. The ultra-hot Neptune LTT9779b is highlighted as the red star. The purple shaded region indicates the Neptunian radius regime, whereas the green shaded region demarcates the “Neptunian desert".
\label{fig:temp_rad}}
\end{figure}

Ultra-hot Neptunes (UHNs) are the Neptunian-mass counterparts to the ultra-hot Jupiters (UHJs), occupying the extreme high-temperature end of the gas-giant population with dayside temperatures $\gtrsim$ 2200K. Theoretically, photoevaporation plays a dominant role during the migration of Neptunian-mass planets to close-in orbits, effectively wearing down their atmospheres and giving rise to the “Neptunian desert" \citep[e.g.,][]{2016A&A...589A..75M, 2018MNRAS.476.5639I}. Despite this, the first UHN deep within the Neptunian desert, LTT9779b, was discovered in 2020 \citep{2020NatAs...4.1148J}—via the exceptional sky coverage of the Transiting Exoplanet Survey Satellite \citep[TESS,][]{2015JATIS...1a4003R}—which remains the only such planet known to date (see Figure \ref{fig:temp_rad}).

The uniqueness of LTT9779b has prompted extensive theoretical speculations and follow-up observations aimed at constraining its formation and evolutionary history, particularly through the characterization of its atmospheric conditions. Initial Spitzer near-IR observations \citep{2020ApJ...903L...6D, 2020ApJ...903L...7C} and high-resolution optical ground-based spectroscopy \citep{2025A&A...695A..26R} have suggested the presence of a potentially high-metallicity atmosphere. However, the unexpectedly high geometric albedo detected in optical observations by CHEOPS \citep{2023A&A...675A..81H, 2025A&A...700A..45S} indicated the possibility of both a metal-rich composition and the presence of dayside clouds—a scenario further supported by JWST NIRISS observations \citep{2024ApJ...962L..20R, 2025NatAs...9..512C}. Most recently, JWST NIRSpec observations \citep{2026AJ....171..215A, Brande2026} have reported the detection of CO and CO$_2$ within a high-metallicity environment.

While previous studies have not provided robust statistical evidence for clouds on LTT9779b, the high geometric albedo could be attributed to either highly reflective dayside clouds or a strongly emitting, high metallicity atmosphere. This degeneracy is particularly pronounced for Neptunian-mass planets; their smaller radii relative to UHJs result in lower signal-to-noise ratios (SNR) and higher observational uncertainties. Specifically, in retrieval frameworks lacking flexible cloud models, the absence of a contributing cloud component may be artificially compensated by an inflated abundance of gaseous species, thereby overestimating the inferred metallicity. While these two scenarios theoretically produce distinct spectral features, the lower SNR associated with LTT9779b’s size can render these differences indistinguishable.

Nonetheless, JWST remains the most sophisticated facility for such characterization. The availability of both NIRISS \citep{2023PASP..135i8001D} and NIRSpec/G395H \citep{2022A&A...661A..80J} observations provides a unique opportunity to statistically probe the presence and nature of dayside clouds on LTT9779b. We focus here on the dayside emission spectrum, as ultra-hot gas giants typically exhibit much stronger features in emission than in transmission, offering a more feasible path toward breaking the cloud-metallicity degeneracy. Recent panchromatic JWST studies have successfully constrained dayside cloud detections in UHJs, such as WASP-121b \citep{2025ApJ...994L..39S} and WASP-19b \citep{ 2026AJ....171..295S}, by leveraging broad wavelength coverage to robustly distinguish the wavelength-dependent contributions of different atmospheric components. However, such characterizations require flexible—and thus computationally intensive—cloud modeling. Despite the complexities introduced by the smaller size of LTT9779b, this study endeavors to statistically constrain its dayside cloud properties, thereby breaking the cloud-metallicity degeneracy and determining the true abundances of the dominant atmospheric gases. In Section \ref{sec:sec2}, we detail the methodologies used in this work, and in Section \ref{sec:sec3}, we present our results and their implication.

\begin{figure*}
\centering
\includegraphics[width=\textwidth]{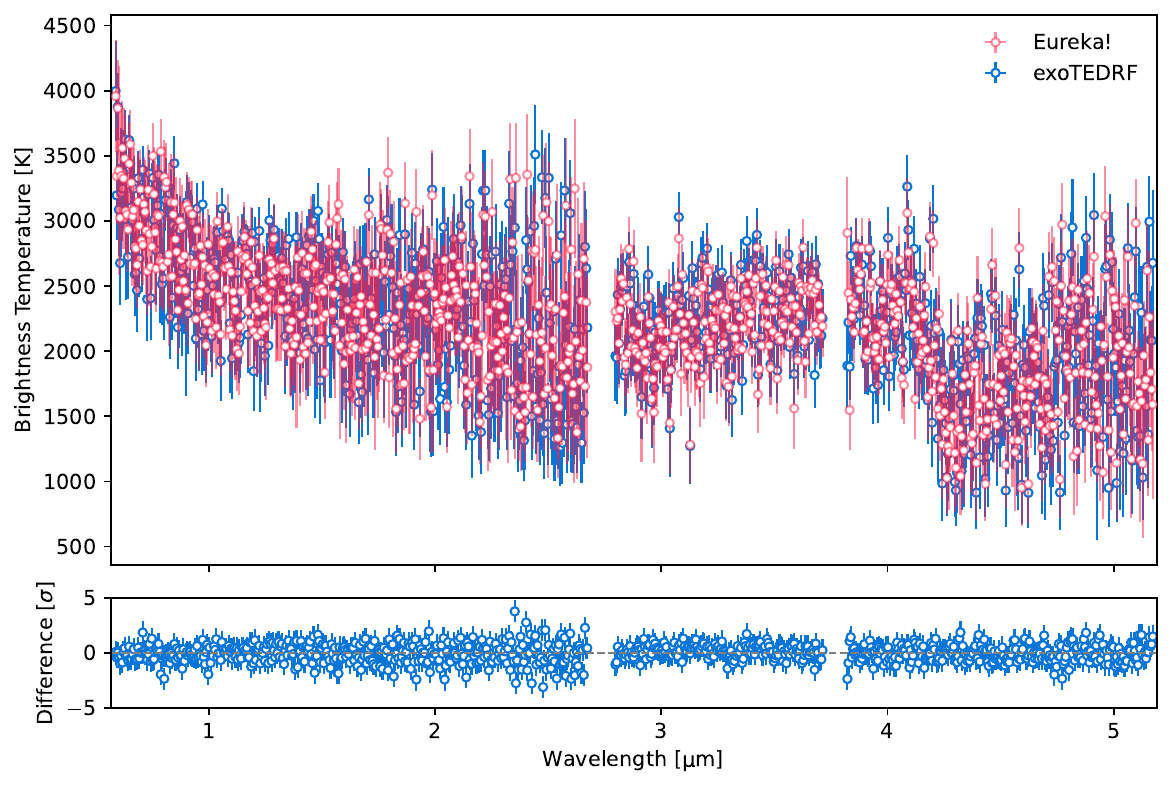}
\caption{Observed panchromatic emission spectra of LTT9779b—combining NIRISS and NIRSpec/G395H observations—shown in terms of brightness temperature (top panel), derived from two independent reductions of the raw data using Eureka! (red) and exoTEDRF (blue), followed by modeling of the binned spectroscopic light curves. The mean absolute difference between the two spectra is $\sim$0.6$\sigma$ (bottom panel), indicating excellent statistical agreement.\label{fig:spectra}}
\end{figure*}

\section{Methodology}\label{sec:sec2}

\subsection{Data reduction and analysis}\label{sec:sec2.1}

JWST conducted phase curve observations of LTT9779b using NIRISS during Cycle 1 (GTO $\#$1201, PI: D. Lafreniere) and NIRSpec/G395H during Cycle 2 (GO $\#$3231, PI: I. Crossfield); both of which encompassed two secondary eclipses. Together, these instrumental setups provide complementary wavelength coverage, delivering medium-resolution spectra spanning $\sim$0.6–5.1 $\mu$m. We accessed the data via Barbara A. Mikulski Archive for Space Telescopes (MAST)\footnote{\url{https://mast.stsci.edu/}} \dataset[(doi: 10.17909/z7wc-qc04)]{\doi{10.17909/z7wc-qc04}}. The NIRISS data comprise 26 segments, of which segments 1–4 and 24–26—covering the two secondary eclipses—were utilized in our analysis. Similarly, the NIRSpec data consist of seven segments, from which we used segments 1 and 7 (covering the secondary eclipses) for both the NRS1 and NRS2 detectors.

Raw data (*uncal.fits) from the JWST Science Calibration Pipeline \citep{2022zndo...7071140B} were processed using two independent, community-developed pipelines: Eureka! \citep{Bell2022} and exoTEDRF \citep[formerly supreme-SPOON,][]{2024JOSS....9.6898R}. Both pipelines are widely adopted and have been rigorously validated across multiple studies (e.g., \citealt{2023Natur.614..664A, 2023NatAs...7.1317L, 2024Natur.626..979P, 2025ApJ...994L..39S, 2026AJ....171..295S} for Eureka!; and \citealt{2023Natur.614..670F, 2024ApJ...962L..20R, 2025ApJ...994L..39S, 2026AJ....171..295S} for exoTEDRF). In addition to the science frames, these pipelines utilize calibration files retrieved from the JWST Calibration Reference Data System (CRDS)\footnote{\url{https://jwst-crds.stsci.edu/}}. The use of two independent pipelines allows for robust cross-validation of our results, a critical step for high-precision atmospheric characterization studies.

The Eureka! pipeline provides a six-stage procedure for reducing and analyzing JWST time-series data, with each stage comprising multiple built-in functions and subprocesses \citep{Bell2022}. For this study, we employed the first three stages—covering calibration, detector-level correction, and optimal spectral extraction. The NIRSpec NRS1 and NRS2 datasets were processed independently. Our reductions largely followed the default configurations specified in the instrument-specific “*.ecf” files for both NIRISS and NIRSpec/G395 spectral reductions, with only minor modifications to optimize performance, as detailed in \cite{2026AJ....171..295S} and \cite{saha_w178}. Similarly, exoTEDRF offers a streamlined framework for raw data reduction, structured into four stages. We utilized the first three stages (calibration, corrections, and spectral extraction) and treated the NIRSpec NRS1 and NRS2 datasets separately. We primarily adhered to the default configurations optimized for NIRISS and NIRSpec/G395H, applying only minor adjustments as outlined in \cite{saha_w178} and \cite{2026AJ....171..295S}.

\subsection{Light curve modeling}\label{sec:sec2.2}

The extracted spectroscopic time-series data (i.e., 2D light curves) from both the Eureka! and exoTEDRF reductions were binned along the wavelength axis to produce the spectroscopic light curves. Given the high SNR of our observations, we tested multiple binning schemes and ultimately adopted a high-resolution approach using uniformly spaced 0.005 $\mu$m bins. This corresponds to a spectral resolution of $\sim$250-2000 across the observed wavelength range. The final binning yielded a total of 873 spectroscopic light curves: 54 from NIRISS Order 2, 364 from NIRISS Order 1, 184 from NIRSpec NRS1, and 271 from NIRSpec NRS2.

The light curves were modeled using our in-house pipeline, ExoELF \citep[ExoplanEts light curves Fitter,][]{2021AJ....162...18S, 2021AJ....162..221S, 2025MNRAS.539..928S, 2025ApJ...994L..39S}, which integrates several widely adopted packages, including batman \citep{2015PASP..127.1161K} for secondary eclipse modeling, dynesty \citep{2020MNRAS.493.3132S} for nested sampling, and celerite \citep{celerite1, celerite2} for Gaussian process (GP) regression. Optimized for high-precision JWST spectroscopy, ExoELF supports flexible detrending strategies and has been validated in several recent JWST-based analyses \citep{2025ApJ...994L..39S, 2026AJ....171..295S, saha_w178}.

We adopted a two-stage procedure to extract the emission spectrum from the light curves. In the first stage, we simultaneously modeled the white light curves from NIRISS Order 1 and NIRSpec NRS1/NRS2, treating orbital parameters and mid-eclipse times as free parameters. NIRISS Order 2 was excluded from this joint fit due to its significantly lower SNR. In the second stage, we fixed the orbital parameters and mid-eclipse times to the values obtained from the white light curve analysis and modeled the spectroscopic light curves, fitting only for the eclipse depth and detrending parameters. After extensively testing various detrending schemes, we adopted a “two-fourth-order" polynomial approach; i.e., each secondary eclipse was modeled with a fourth-order polynomial \citep{2025ApJ...994L..39S}. This configuration effectively detrends both eclipses in each light curve with minimal parameter overhead while maintaining a likelihood comparable to more complex models. To ensure numerical stability, we used a fourth-order Legendre polynomial expansion instead of a standard polynomial, as its orthogonality mitigates correlations between higher-order terms. All joint light curve fittings were performed using nested sampling via dynesty.

The emission spectra obtained from both independent Eureka! and exoTEDRF pipelines show a mean absolute difference of only $\sim$0.6$\sigma$ (see Figure \ref{fig:spectra}), demonstrating excellent statistical agreement. However, consistent with our previous findings \citep{2025ApJ...994L..39S, saha_w178}, we noted that the exoTEDRF spectrum exhibits notably smaller estimated uncertainties, despite lacking a corresponding reduction in variance relative to the Eureka! spectrum. Consequently, we adopt the Eureka! spectrum as our default dataset for atmospheric retrieval analyses, while performing parallel analyses on both datasets to cross-validate the robustness of our results.

\subsection{Atmospheric retrieval analyses}\label{sec:sec2.3}

To characterize the dayside atmospheric properties of LTT9779b, we performed a series of atmospheric retrieval analyses. While retrievals are often computationally intensive and susceptible to parameter degeneracies when datasets are sparse, the available panchromatic wavelength coverage—combined with high SNR and spectral resolution—presents a robust opportunity to infer well-constrained atmospheric conditions. For the forward modeling, we employed petitRADTRANS \citep[pRT,][]{2019A&A...627A..67M}, a widely adopted and validated radiative transfer code within the exoplanet community \citep[e.g.,][]{2024ApJ...973L..41I, 2024MNRAS.527.7079P, 2024A&A...690A..63B, 2025ApJ...994L..39S, 2026AJ....171..295S}. pRT utilizes a one-dimensional radiative transfer formalism that provides the comprehensive functionality necessary for complex retrievals while maintaining the computational efficiency required to interpret data-intensive observations, such as those from JWST.

Our retrieval framework is based on the pRT retrieval routines \citep{2024JOSS....9.5875N}, which incorporates PyMultinest \citep{2009MNRAS.398.1601F, 2014A&A...564A.125B} for nested sampling of the posterior space. Given the sufficient spectral resolution of our extracted emission spectra, we utilized the line-by-line opacity mode with a model resolution of R = 20,000, providing the necessary precision for a robust interpretation of our datasets. To account for potential observational offsets between the NIRISS, NIRSpec/NRS1, and NIRSpec/NRS2 data, we included scaling factors for NIRISS ($f_{\mathrm{s, NIRISS}}$) and NIRSpec/NRS2 ($f_{\mathrm{s, NRS2}}$) as free parameters in the retrievals (see Table \ref{tab:retrieval_priors}).

We adopted the analytical pressure-temperature (P–T) profile formalism of \citet{2010A&A...520A..27G} (hereafter, Guillot), which provides sufficient flexibility for robust interpretations without introducing unnecessary computational complexity. This parametrization is defined by four variables: the equilibrium (irradiation) temperature (T$_{\mathrm{eq}}$), the internal temperature (T$_{\mathrm{int}}$), the visible-to-infrared opacity ratio ($\gamma$), and the mean infrared opacity ($\kappa_{\mathrm{IR}}$).

Additionally, to assess the robustness of our results with respect to this formalism, we also introduced an highly flexible P–T parametrization based on the simpler framework of \citet{2009ApJ...707...24M}. We refer to this as the neo-MS parametrization, which follows the general expression: 

\begin{equation}
\text{T}(\text{P}) = \text{T}_{i-1} + \text{sgn}(\gamma_i) \left| \gamma_i \ln \left( \frac{\text{P}}{\text{P}_{i-1}} \right) \right|^{\xi_i}
\end{equation}

Here $\text{sgn}(\gamma_i)$ denotes the sign of the temperature gradient, $\gamma_i$ sets the gradient scale, and $\xi_i$ controls the power-law curvature exponent. To ensure a continuity in a multi-layer configuration, the boundary temperatures (T$_{i}$) are recursively linked to the parameters of the overlying layers (T$_{i-1}$ and P$_{i-1}$), with T$_{0}$ serving as the independent anchor temperature at the top of the atmosphere (P$_{0}$).

In our implementation, we adopted a four-layer model extending from the top-of-atmosphere pressure P$_{0}$ through transition boundaries at P$_{1}$, P$_{2}$, and P$_{3}$:

\begin{equation}
\text{T}(\text{P}) = 
\begin{cases} 
\text{T}_0 + \text{sgn}(\gamma_1) \left| \gamma_1 \ln \left( \frac{\text{P}}{\text{P}_0} \right) \right|^{\xi_1} & \text{if } \text{P}_0 \leq \text{P} \leq \text{P}_1 \\
\text{T}_1 + \text{sgn}(\gamma_2) \left| \gamma_2 \ln \left( \frac{\text{P}}{\text{P}_1} \right) \right|^{\xi_2} & \text{if } \text{P}_1 < \text{P} \leq \text{P}_2 \\
\text{T}_2 + \text{sgn}(\gamma_3) \left| \gamma_3 \ln \left( \frac{\text{P}}{\text{P}_2} \right) \right|^{\xi_3} & \text{if } \text{P}_2 < \text{P} \leq \text{P}_3 \\
\text{T}_3 & \text{if } \text{P} > \text{P}_3
\end{cases}
\end{equation}

To enforce the physical requirement that $\text{P}_1 < \text{P}_2 < \text{P}_3$ throughout the retrieval, we do not sample the transition pressures directly. Instead, we define the pressure boundaries using dimensionless coordinate factors, $\delta_{\mathrm{P}_{i}}$, which partition the log-pressure space between the minimum and maximum atmospheric boundaries. The transition pressures are recursively computed as:

\begin{equation}
\begin{aligned}
\log \text{P}_1 &= \log \text{P}_{\mathrm{min}} + (\log \text{P}_{\mathrm{max}} - \log \text{P}_{\mathrm{min}}) \cdot \delta_{\mathrm{P}_{1}} \\
\log \text{P}_2 &= \log \text{P}_1 + (\log \text{P}_{\mathrm{max}} - \log \text{P}_1) \cdot \delta_{\mathrm{P}_{2}} \\
\log \text{P}_3 &= \log \text{P}_2 + (\log \text{P}_{\mathrm{max}} - \log \text{P}_2) \cdot \delta_{\mathrm{P}_{3}}
\end{aligned}
\end{equation}

We employed both free and equilibrium chemistry approaches to characterize the atmospheric abundances. In the free chemistry approach, the abundances—expressed as log-mass mixing ratios (log-MMRs)—were treated as vertically uniform free parameters, with priors typically ranging between -14 to 0 (see Table \ref{tab:retrieval_priors}). To improve computational efficiency, the upper bounds for species showing no evidence of high abundance in preliminary fits were lowered. For the equilibrium chemistry models, we utilized precomputed abundance tables provided by pRT, assuming chemical equilibrium at each pressure level.

We included the opacities of all key molecules expected to contribute significantly to the emission spectra, including H$_2$O \citep{2018MNRAS.480.2597P}, CO \citep{2010JQSRT.111.2139R}, CO$_2$ \citep{2010JQSRT.111.2139R}, SiO \citep{2013MNRAS.434.1469B}, CH$_4$ \citep{2020ApJS..247...55H}, HCN \citep{2006MNRAS.367..400H}, C$_2$H$_2$ \citep{2013JQSRT.130....4R}, PH$_3$ \citep{2015MNRAS.446.2337S}, H$_2$S \citep{2013JQSRT.130....4R}, TiO \citep{2019A&A...627A..67M}, and VO \citep{2019A&A...627A..67M}. Rayleigh scattering by H$_2$ \citep{1962ApJ...136..690D} and He \citep{1965PPS....85..227C} was accounted for, alongside collision-induced absorption (CIA) from H$_2$-H$_2$ \citep{2001JQSRT..68..235B, 2002A&A...390..779B} and H$_2$-He \citep{1988ApJ...326..509B, 1989ApJ...336..495B}.

Furthermore, we included several condensate species expected to form clouds at these temperatures, following the formalism of \citet{2001ApJ...556..872A}. These include Fe(s) \citep{2018MNRAS.475...94K, 1994ApJ...421..615P, 1997A&A...327..743H, 2001ApJ...556..872A}, MgSiO$_3$(s) \citep{1994A&A...292..641J, 2001ApJ...556..872A}, Mg$_2$SiO$_4$(s) \citep{1973PSSBR..55..677S, 2017Icar..289...42C}, Al$_2$O$_3$(s) \citep{1995Icar..114..203K, Stull1947}, SiO$_2$(s) \citep{2018MNRAS.475...94K, 1997A&A...327..743H, 1995Icar..114..203K, Stull1947}, TiO$_2$(s) \citep{2018MNRAS.475...94K, 2011A&A...526A..68Z, 2003ApJS..149..437P, 2016arXiv160704866S, 2019RJPCA..93.1024S}, and CaTiO$_3$(s) \citep{2018MNRAS.475...94K, 2003ApJS..149..437P, 1998JPCM...10.3669U, 2019RJPCA..93.1024S}. We have considered the opacities of all condensates in their crystalline solid phases. The abundances of these species at the cloud base were treated as free parameters in both the free and equilibrium chemistry models. Additionally, the sedimentation efficiency  (f$_{\mathrm{sed}}$), log-normal width of particle size distribution ($\sigma$), the vertical eddy diffusion coefficient (K$_{\mathrm{zz}}$), and the cloud fraction (f$_{\mathrm{c}}$) were treated as free parameters to collectively constrain the cloud profiles \citep{2001ApJ...556..872A}. We also accounted for the dayside scattering of starlight using a dayside-averaged irradiance.

\begin{figure*}
\centering
\includegraphics[width=\textwidth]{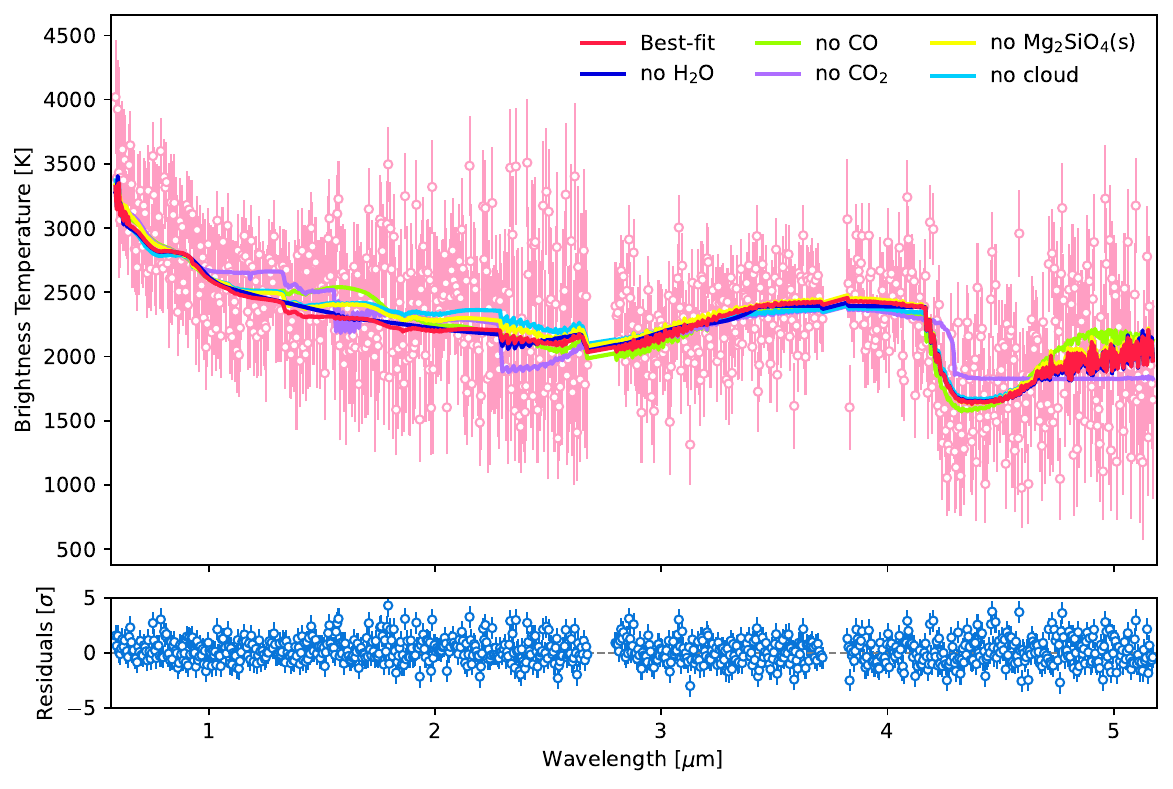}
\caption{Observed panchromatic emission spectra (Eureka!) shown with the retrieved best-fit free chemistry model (with Guillot P–T parametrization) and associated residuals. Additional models—excluding specific gaseous species and cloud components—are also overplotted for comparison.
\label{fig:modelplot}}
\end{figure*}

\begin{table*}
	\centering
	\caption{Molecular abundances (in log-MMRs) and log-evidence from the free chemistry atmospheric retrievals.}
	\label{tab:tab1}
	\begin{tabular}{lccccccc}
		\hline
		Model & [H$_2$O] & [CO] & [CO$_2$] & [SiO] & [TiO] & [VO] & $\ln Z$ \\
		\hline
		Eureka! (Guillot)\\
		Best-fit & $-4.83_{-0.26}^{+0.23}$ & $-2.76_{-0.45}^{+0.53}$ & $-5.06_{-0.26}^{+0.30}$ & $<-6.51$ & $<-11.87$ & $<-11.32$ & 32456.29 \\
		No H$_2$O & – & $-2.42_{-0.35}^{+0.37}$ & $-4.9_{-0.22}^{+0.22}$ & $<-6.25$ & $<-11.94$ & $<-11.35$ & 32455.11 \\
        No CO & $-4.83_{-0.14}^{+0.13}$ & – & $-5.32_{-0.13}^{+0.16}$ & $<-7.25$ & $<-11.73$ & $<-11.18$ & 32444.36 \\
        No CO$_2$ & $-4.9_{-0.17}^{+0.19}$ & $>-0.28$ & – & $<-3.55$ & $<-11.79$ & $<-11.27$ & 32417.93 \\
        No SiO & $-4.82_{-0.26}^{+0.23}$ & $-2.78_{-0.44}^{+0.5}$ & $-5.09_{-0.25}^{+0.29}$ & – & $<-11.96$ & $<-11.36$ & 32455.61 \\
        No TiO & $-4.82_{-0.28}^{+0.25}$ & $-2.69_{-0.42}^{+0.52}$ & $-5.04_{-0.23}^{+0.29}$ & $<-6.51$ & – & $<-11.46$ & 32456.61 \\
        No VO & $-4.83_{-0.27}^{+0.23}$ & $-2.7_{-0.41}^{+0.49}$ & $-5.04_{-0.24}^{+0.26}$ & $<-6.38$ & $<-11.95$ & – & 32456.32 \\
        No Mg$_2$SiO$_4$(s) & $-4.74_{-0.16}^{+0.20}$ & $-3.08_{-0.37}^{+0.61}$ & $-5.23_{-0.23}^{+0.35}$ & $<-6.69$ & $<-11.52$ & $<-11.13$ & 32454.59 \\
        No clouds & $-4.50_{-0.36}^{+0.66}$ & $>-3.03$ & $-4.77_{-0.45}^{+0.75}$ & $<-6.08$ & $<-11.76$ & $<-11.19$ & 32452.27 \\
        No clouds (CO limit) & $-4.68_{-0.26}^{+0.25}$ & $-2.73_{-0.45}^{+0.51}$ & $-5.06_{-0.26}^{+0.31}$ & $<-6.34$ & $<-11.87$ & $<-11.29$ & 32451.59 \\
		\hline
        Eureka! (neo-MS)\\
        Best-fit & $-4.82_{-0.18}^{+0.17}$ & $-2.66_{-0.40}^{+0.49}$ & $-5.01_{-0.21}^{+0.25}$ & $<-6.69$ & $<-11.75$ & $<-11.49$ & 32453.99 \\
        No H$_2$O & – & $-1.91_{-0.30}^{+0.32}$ & $-4.58_{-0.15}^{+0.16}$ & $<-6.59$ & $<-11.63$ & $<-11.10$ & 32443.04 \\
        No CO & $-4.66_{-0.16}^{+0.15}$ & – & $-4.76_{-0.18}^{+0.22}$ & $<-6.67$ & $<-11.57$ & $<-11.19$ & 32427.82 \\
        No CO$_2$ & $-4.93_{-0.28}^{+0.20}$ & $>-0.30$ & – & $<-4.23$ & $<-11.52$ & $<-11.17$ & 32399.61 \\
        No SiO & $-4.83_{-0.17}^{+0.17}$ & $-2.78_{-0.34}^{+0.40}$ & $-5.09_{-0.19}^{+0.23}$ & – & $<-11.69$ & $<-11.40$ & 32453.05 \\
        No Mg$_2$SiO$_4$(s) & $-4.82_{-0.18}^{+0.17}$ & $-2.95_{-0.39}^{+0.56}$ & $-5.16_{-0.22}^{+0.30}$ & $<-6.85$ & $<-11.80$ & $<-11.35$ & 32451.55 \\
        No clouds & $-4.53_{-0.23}^{+0.21}$ & $>-0.87$ & $-3.86_{-0.27}^{+0.26}$ & $<-6.54$ & $<-11.77$ & $<-11.29$ & 32445.26 \\
        \hline
		exoTEDRF (Guillot)\\
		Best-fit & $<-5.75$ & $-3.21_{-0.24}^{+0.26}$ & $-5.42_{-0.14}^{+0.18}$ & $<-6.66$ & $<-11.95$ & $<-11.31$ & 32273.00 \\
        No H$_2$O & – & $-3.57_{-0.19}^{+0.33}$ & $-5.56_{-0.13}^{+0.19}$ & $<-6.93$ & $<-11.98$ & $<-11.39$ & 32273.04 \\
        No CO & $-5.87_{-1.08}^{+0.32}$ & – & $-5.40_{-0.10}^{+0.10}$ & $<-7.36$ & $<-11.58$ & $<-11.21$ & 32254.40 \\
        No CO$_2$ & $<-5.83$ & $>-0.24$ & – & $<-7.07$ & $<-11.34$ & $<-11.02$ & 32236.38 \\
        No Mg$_2$SiO$_4$(s) & $-5.61_{-0.55}^{+0.26}$ & $-3.52_{-0.17}^{+0.18}$ & $-5.58_{-0.10}^{+0.12}$ & $<-6.81$ & $<-11.43$ & $<-10.26$ & 32267.78 \\
        No clouds & $<-5.46$ & $-3.37_{-0.23}^{+0.26}$ & $-5.48_{-0.13}^{+0.17}$ & $<-6.70$ & $<-11.80$ & $<-11.29$ & 32260.19 \\
		\hline
	\end{tabular}
\end{table*}

\section{Results and Discussion}\label{sec:sec3}

From the free chemistry retrievals of the Eureka! spectrum using the Guillot P–T profile, we obtained precise constraints on the abundances of H$_2$O, CO, and CO$_2$ in the dayside atmosphere of LTT9779b, along with tentative upper limits for SiO, TiO, and VO (see Table \ref{tab:tab1}, Figure \ref{fig:cor_eu}). To assess the statistical significance \citep{Kass01061995, 2008ConPh..49...71T} of each molecular detection, we performed Bayesian model comparisons by conducting “leave-one-out" retrievals, in which individual species were excluded (see Table \ref{tab:tab1}, Figure \ref{fig:modelplot}). These comparisons yield robust detections of CO (4.88$\sigma$, $B_{10} = 1.51 \times 10^5$) and CO$_2$ (8.76$\sigma$, $B_{10} = 4.55 \times 10^{16}$) (see Table \ref{tab:tab1}, Figure \ref{fig:modelplot}). We also find weak evidence for H$_2$O (1.53$\sigma$, $B_{10} = 3.24$), while SiO, TiO, and VO show no statistically significant detections. Although strong upper limits are placed on these species, the current S/N of the data remains insufficient for definitive detection.

\begin{figure*}
\centering
\includegraphics[width=\textwidth]{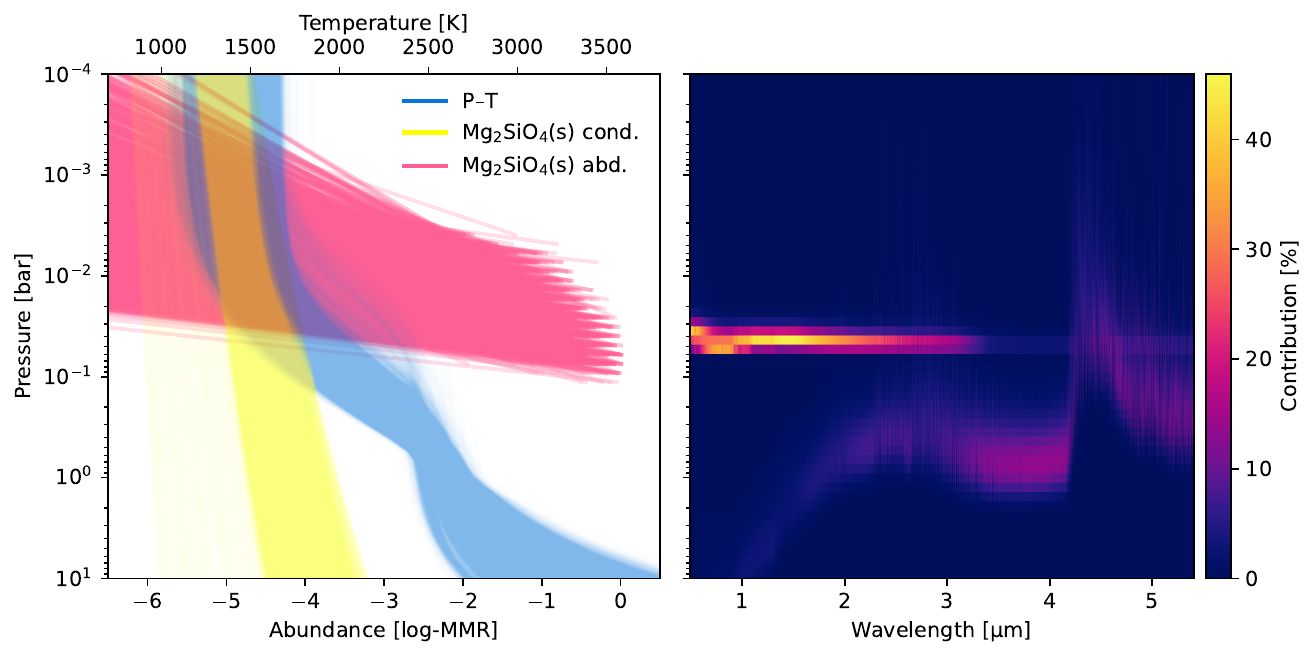}
\includegraphics[width=\textwidth]{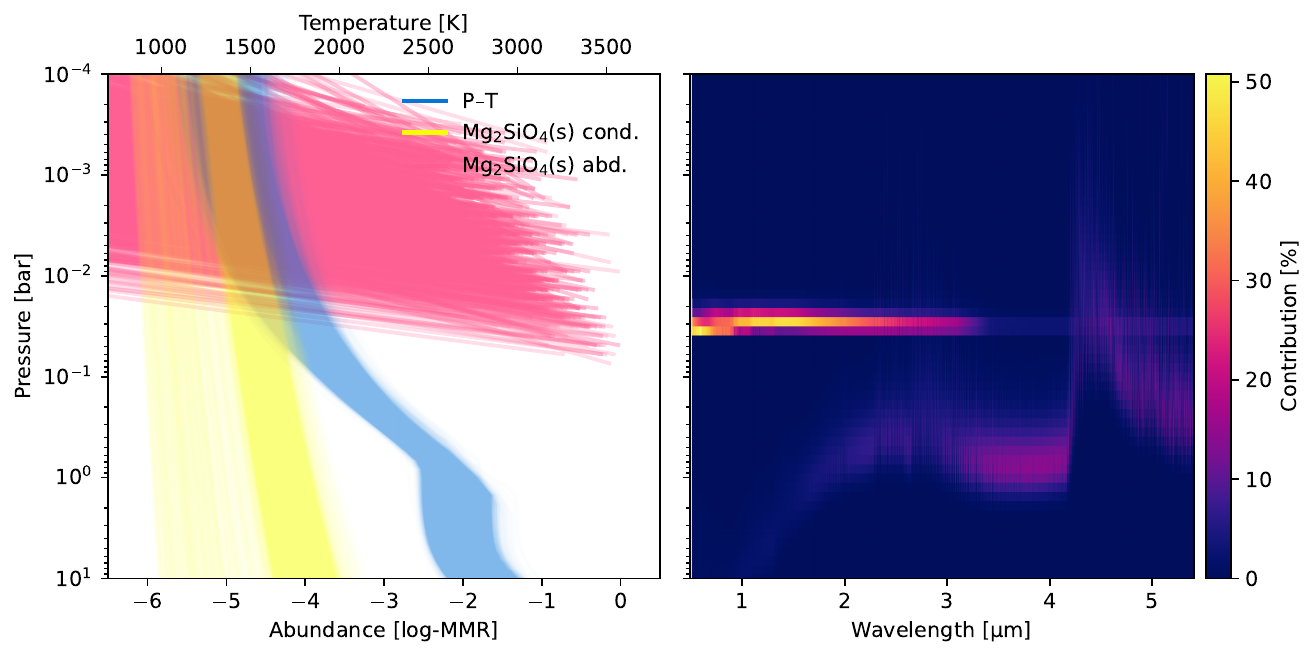}
\caption{Retrieved posterior P–T profiles from the free chemistry retrievals of the Eureka! spectrum using the Guillot P–T parametrization (top) and the neo-MS P–T parametrization (bottom), together with the corresponding posteriors for Mg$_2$SiO$_4$(s) condensation curves and abundance profiles. The median contribution functions associated with these retrievals are also shown.
\label{fig:contribution}}
\end{figure*}

The retrievals also yielded strong constraints on the cloud properties and the base-abundance for Mg$_2$SiO$_4$(s), indicating its presence in the dayside atmosphere. Contribution functions show that the emission spectrum is dominated by strong dayside reflection from the cloud deck at shorter wavelengths, while molecular features from CO and CO$_2$ contribute significantly to the flux at longer wavelengths (see Figure \ref{fig:contribution}). This atmospheric configuration provides a robust explanation for the exceptionally high geometric albedo of LTT9779b and is consistent with previous predictions following CHEOPS and Spitzer observations \citep{2025A&A...700A..45S}.

\begin{figure*}
\centering
\includegraphics[width=\textwidth]{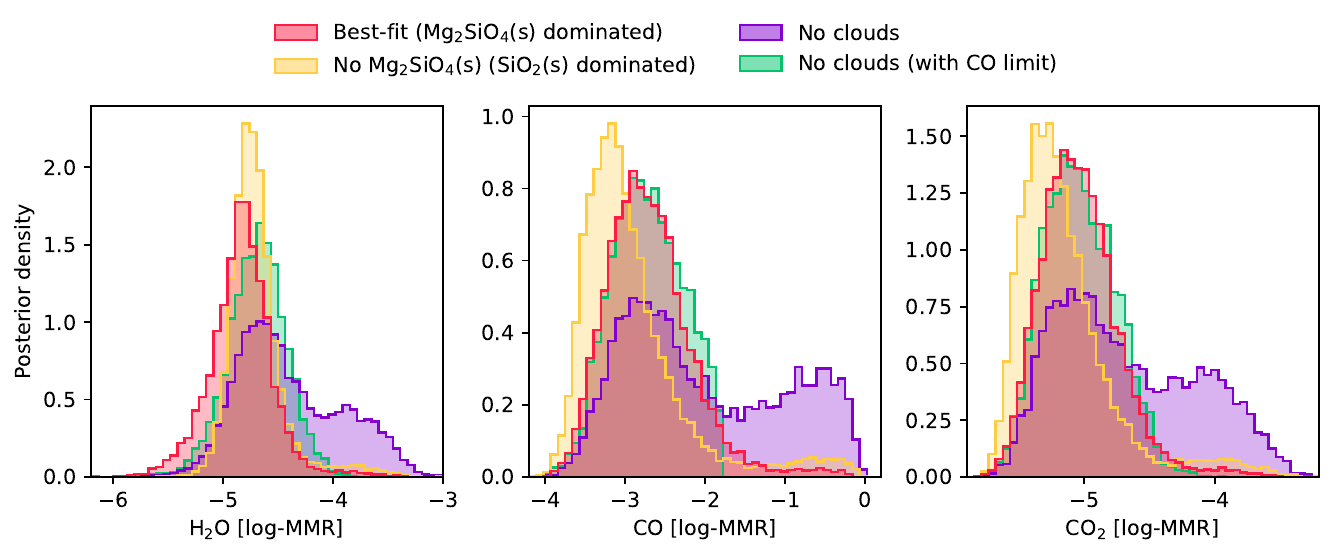}
\includegraphics[width=\textwidth]{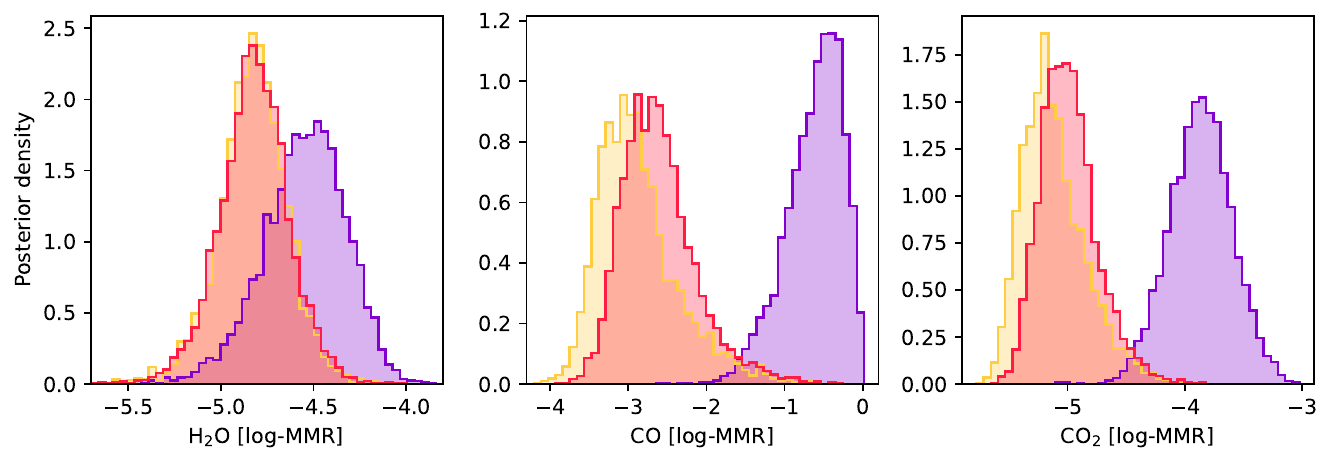}
\caption{Comparison of abundance posteriors for key molecular species—H$_2$O, CO, and CO$_2$—from free chemistry retrievals of the Eureka! spectrum, using the Guillot P–T parametrization (top panels) and the neo-MS P–T parametrization (bottom panels), across different model conditions.
\label{fig:comparison}}
\end{figure*}

To assess the robustness of the cloud detection, we again performed Bayesian model comparisons, which yield a statistical preference of 2.84$\sigma$ ($B_{10} = 55.71$) for the cloudy model over a cloud-free alternative (see Table \ref{tab:tab1}, Figure \ref{fig:modelplot}). However, a comparison of the inferred molecular abundances between these models shows that the cloud-free scenario artificially inflates the abundances of dominant species—H$_2$O, CO, and CO$_2$—to produce the higher flux at shorter wavelengths, while simultaneously maintaining lower posterior abundance distributions to fit the molecular features at longer wavelengths (see Figure \ref{fig:comparison}). This bifurcation in abundance constraints highlights the physical necessity of including clouds to explain the observed spectrum. Furthermore, these inflated abundances are effectively bounded by the upper limit of the CO prior—i.e., a 100\% CO atmosphere—which is physically implausible. Such an unphysically high CO abundance, in the absence of proper cloud contributions, drives the atmospheric metallicity towards infinity ([M/H] $\rightarrow$ $\infty$), thereby explaining the highly metal-enriched atmospheric scenarios inferred in previous studies \citep[e.g.,][]{2025A&A...700A..45S, 2025NatAs...9..512C, 2026AJ....171..215A, Brande2026}.

To mitigate the effect of these non-physical molecular abundances in the cloud-free model, we performed additional retrievals with the CO abundance capped at –1.8 (log-MMR) (see Figure \ref{fig:comparison}, Table \ref{tab:tab1}). This resulted in stronger evidence for the presence of clouds, at 3.07$\sigma$ ($B_{10} = 110.26$). To evaluate the specific requirement for Mg$_2$SiO$_4$(s) clouds, we performed retrievals excluding this species, which instead favor a SiO$_2$(s) cloud deck; however, this solution is less significant by 1.84$\sigma$ ($B_{10} = 5.48$) compared to the Mg$_2$SiO$_4$(s) case. Moreover, the SiO$_2$(s) dominated model appeared to overfit the molecular features, leading to slightly lower abundances of CO and CO$_2$.

\begin{figure*}
\includegraphics[width=0.525\textwidth]{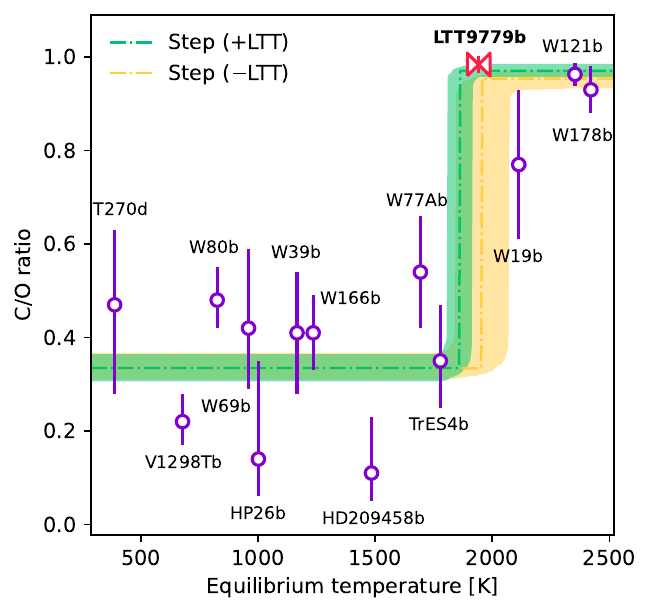}
\hspace{-0.02\textwidth}
\includegraphics[width=0.475\textwidth]{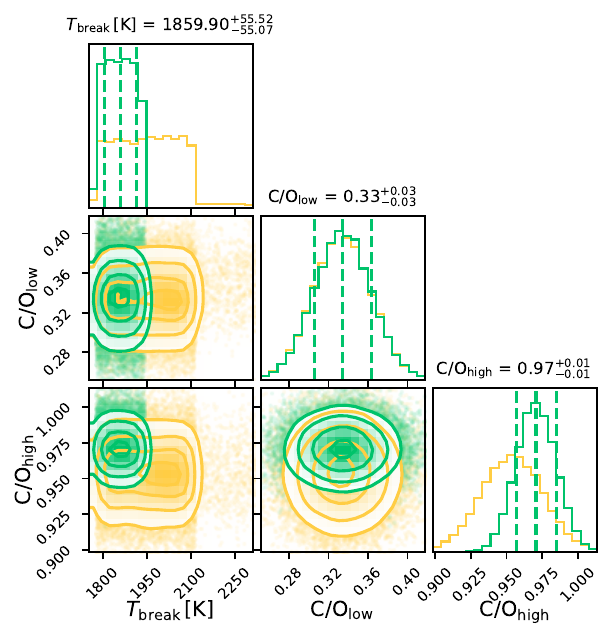}
\caption{Estimated C/O ratio of LTT9779b (red bowtie) shown alongside well-constrained JWST literature values for other exoplanets spanning a wide range of equilibrium temperatures (left panel). Step function fits, illustrating the onset of super-solar C/O ratios in ultra-hot gas giants, are shown both including and excluding LTT9779b (median and 1$\sigma$ confidence intervals), along with the corresponding fit posteriors (right panel).
\label{fig:c_o}}
\end{figure*}

Next, using the neo-MS P–T profile in the free chemistry retrievals of the Eureka! spectrum, we obtained consistent, albeit slightly tighter, constraints on the abundances of H$_2$O, CO, and CO$_2$, along with upper limits for SiO, TiO, and VO (see Table \ref{tab:tab1}, Figure \ref{fig:m_cor_eu}). Additionally, “leave-one-out" retrievals confirm robust detections of CO (7.24$\sigma$, $B_{10} = 2.32 \times 10^{11}$) and CO$_2$ (10.43$\sigma$, $B_{10} = 4.15 \times 10^{23}$). We also find a much stronger evidence for H$_2$O at 4.68$\sigma$ ($B_{10} = 5.71 \times 10^4$), while SiO, TiO, and VO show no statistically significant detections. The tighter abundance constraints and higher statistical significances obtained here can be attributed to the greater flexibility of the neo-MS P–T parametrization relative to the Guillot profile. This increased flexibility provides the best-fit model with additional freedom to converge more directly towards the higher-likelihood solution, while also allowing the deficient models to deviate more strongly from the best-fit scenario.

In addition, these retrievals yield stronger evidence for the presence of clouds at 4.18$\sigma$ ($B_{10} = 6.15 \times 10^3$), with a specific preference for Mg$_2$SiO$_4$(s) clouds at 2.21$\sigma$ ($B_{10} = 11.39$). Comparing the posterior distributions of key molecular abundances between cloudy and cloud-free models, we find that the cloud-free scenario exhibits even stronger abundance enhancements, with the posteriors becoming entirely bounded by the upper limit of the CO prior in this case (see Figure \ref{fig:comparison}). Such stronger deviations from the best-fit scenario are likewise driven by the increased flexibility of the P–T parametrization, further reinforcing the necessity of clouds to explain the observations. These comparisons also highlight the crucial role of P–T parametrizations in accurately inferring molecular abundances from high S/N observations, such as those obtained with JWST.

Inferring from the posterior distributions of the cloud properties obtained from these retrievals, the overall large f$_{\mathrm{sed}}$ indicates that the cloud decks are vertically less extended and consist of relatively large particles (see Figures \ref{fig:contribution}, \ref{fig:cor_eu}, and \ref{fig:m_cor_eu}). The high reflectivity may therefore be interpreted as arising from a combination of an optically thick cloud deck and the wavelength-dependent scattering properties of the condensate species. In contrast, an optically thin cloud deck would likely remain undetectable at the current S/N levels achievable in exoplanet atmospheric observations. The inferred cloud-fraction parameters also indicate highly patchy cloud formation. As suggested by the contribution function, the majority of the CO/CO$_2$ spectral features originate from atmospheric layers deeper than the cloud deck, implying that these features are primarily observed through the gaps between the cloud patches.

To further assess the robustness of these findings from the Eureka! spectrum, we performed analogous retrievals on the exoTEDRF spectrum using the Guillot P–T profile. While these retrievals yielded more tightly constrained abundances for CO and CO$_2$ compared to the Eureka! results, the H$_2$O abundance remained poorly constrained towards its lower bound (see Table \ref{tab:tab1}, Figure \ref{fig:cor_et}). For SiO, TiO, and VO, we again obtained well-constrained upper limits. Overall, the inferred abundances remained broadly consistent between the two reduction pipelines. Using “leave-one-out" retrievals, we confirmed robust detections of CO (6.1$\sigma$, $B_{10} = 1.19 \times 10^8$) and CO$_2$ (8.56$\sigma$, $B_{10} = 8.00 \times 10^{15}$) (see Table \ref{tab:tab1}); however, we find no significant evidence for H$_2$O, nor for SiO, TiO, or VO. The tighter abundance constraints and higher statistical significances are direct consequences of the lower uncertainties in the light curves reduced with exoTEDRF, as previously discussed.

Furthermore, the exoTEDRF retrievals yielded a 5.06$\sigma$ ($B_{10} = 3.64 \times 10^5$) significance for the presence of clouds, with a specific preference for Mg$_2$SiO$_4$(s) clouds at 3.23$\sigma$ ($B_{10} = 185.16$) (see Figure \ref{fig:contribution_et}). Interestingly, the bifurcation of molecular abundances within the cloud-free model is less pronounced in this case (see Figure \ref{fig:comparison_exotedrf}). This is because the lower uncertainties in the exoTEDRF spectrum forced the model to more closely reproduce the shallow molecular features across the wavelength range. However, the small bifurcated CO posterior density accumulating near the upper prior limit suggests that unphysically high metallicities would still be inferred if a higher—and physically unrealistic—upper bound were adopted.

While the definitive identification of cloud species remains challenging, the limitations in this study primarily arise from the smaller radius of LTT9779b relative to Jovian planets (e.g., WASP-121b \citep{2025ApJ...994L..39S}, and WASP-19b \cite{2026AJ....171..295S}), which results in smaller atmospheric signals. Nonetheless, the capabilities of JWST demonstrated here are remarkable, as such minute features are now becoming detectable. Further follow-up observations of LTT9779b could potentially improve these detection significances and enable the definitive detection of refractory species such as SiO, TiO, and VO in this atmosphere.

To further investigate the atmospheric state, we performed equilibrium chemistry retrievals (using the Guillot P–T profile), which failed to provide an adequate fit to the observed spectrum. Specifically, the expected abundances of H$_2$O, SiO, TiO, and VO (and to some extent CO) in an equilibrium chemistry description were higher than those obtained from the free-chemistry retrievals. The model could only fit the CO$_2$ abundance—the strongest molecular contributor to the observed spectral features—with good accuracy. However, the expected abundances of other molecules for the corresponding T–P conditions were higher than the observed spectrum suggested. This indicates strong deviations from chemical equilibrium in the contributing atmospheric layers, primarily driven by condensation and oxygen depletion.

Using the estimated free-chemistry molecular abundances from the Eureka! spectrum with the Guillot P–T profile, we derive a C/O ratio of 0.984 $\pm$ 0.019, which is an extremely high super-solar value. Retrievals using the neo-MS P–T profile yield a consistent estimate of 0.987 $\pm$ 0.014. Although these uncertainties appear exceptionally small, they primarily arise from the significant contrast between the inferred abundance of the dominant carbon- and oxygen-bearing species (CO) and those of other oxygen-bearing species (H$_2$O and CO$_2$). This disparity effectively drives the C/O ratio towards $\sim$1; consequently, the narrow uncertainty is not necessarily representative of the true uncertainties in the inferred molecular abundances. Because the exoTEDRF retrievals did not tightly constrain the H$_2$O abundance, we instead adopted its 1$\sigma$ upper limit, yielding a C/O ratio of 0.992 $\pm0.005$. As noted previously, this uncertainty is likewise strongly influenced by the high inferred CO abundance, although the resulting C/O ratio remains consistent with that derived from the Eureka! spectrum. 

Notably, the super-solar C/O ratio of LTT9779b aligns with recent findings across several UHJs. To analyze broader trends in observed C/O ratios relative to the equilibrium temperature, we compiled a sample of precisely measured values from the literature (Figure~\ref{fig:c_o}). These values and their corresponding equilibrium temperatures were adopted from several recent studies \citep{2024arXiv240303325B, 2024AJ....168...14W, 2024ApJ...963L...5X, 2024ApJ...962L..30E, 2011Natur.469...64M, 2024A&A...685A..64K, 2025arXiv250601800W, 2025MNRAS.539.1381M, 2025AJ....170..292G, 2025arXiv251027223D, 2025AJ....170..165B, 2025AJ....170...50M, 2025ApJ...994L..39S, 2026AJ....171..295S, saha_w178, 2023ApJS..268....2S, 2024ApJS..274...13S}. Within this population, LTT9779b occupies a unique parameter space, exhibiting one of the highest known C/O ratios at an equilibrium temperature marking the lower boundary of the ultra-hot gas-giant regime.

To assess the emerging trend of super-solar C/O ratios among ultra-hot gas giants and determine if LTT9779b follows this pattern, we modeled the compiled C/O ratios using a step function fit, consisting of low- and high-temperature C/O plateaus ($C/O_{\rm low}$ and $C/O_{\rm high}$, respectively) separated by a transition temperature ($T_{\rm break}$), both with and without the inclusion of LTT9779b (see Figure \ref{fig:c_o}). Including LTT9779b in the sample yielded $C/O_{\text{low}} = 0.334_{-0.029}^{+0.029}$, $C/O_{\text{high}} = 0.971_{-0.014}^{+0.014}$, and $T_{\text{break}} = 1859.9_{-55.1}^{+55.5}$ K. Conversely, excluding LTT9779b resuled in $C/O_{\text{low}} = 0.334_{-0.029}^{+0.030}$, $C/O_{\text{high}} = 0.954_{-0.022}^{+0.021}$, and $T_{\text{break}} = 1956.2_{-120.7}^{+115.7}$ (see Figure \ref{fig:c_o}). These results show that the model including LTT9779b is fully consistent with the fit obtained when the planet is excluded, while modestly improving the overall fit ($\Delta$BIC $\sim$ 1.5). This suggests that the high C/O ratio inferred for LTT9779b likely shares a common origin with those observed in other UHJs.

We emphasize that a larger sample of precisely estimated C/O ratios is required to validate this apparent trend among ultra-hot gas giants. One potential driver of this trend may be the sequestration of oxygen-bearing refractory condensates—such as Mg$_2$SiO$_4$(s) and CaTiO$_3$(s)—on the cooler nightside of these planets. Their immense dayside temperatures act as a catalyst for this process, as refractory species such as SiO and TiO could exist in gas form. These species form refractory condensates higher in the atmosphere, which are subsequently transported to the nightside and “cold-trapped", thereby reducing the dayside oxygen budget \citep{2025ApJ...994L..39S, 2026AJ....171..295S, 2023MNRAS.520.4683F, 2024ApJ...963...67C}. The estimated super-solar C/O ratio of LTT9779b strengthens this hypothesis, adding the first and only known UHN to the sample. The oxygen depletion is further corroborated by the lower abundances of oxygen-bearing molecules such as CO, H$_2$O, and SiO discussed previously. Consequently, this sequestration leads to a slightly sub-solar metallicity for the dayside atmosphere when inferred from observed abundances.

Our findings establish key atmospheric benchmarks for the UHN exoplanet LTT9779b, which will help advance models of atmospheric evolution for ultra-hot gas-giants and Neptunian desert objects. Finally, we emphasize the necessity of high S/N panchromatic observations for making physically meaningful atmospheric inferences. However, achieving the requisite S/N levels may require multiple JWST transits or secondary eclipse follow-ups for several of these targets.\\

We thank the anonymous reviewer for their helpful comments and suggestions on this manuscript. SS acknowledges the ANID BASAL project FB210003, Centro de Excelencia en Astrofísica y Tecnologías Afines (CATA), to support this research. The computations presented in this work were performed using the Geryon-3 supercomputing cluster, which was assembled and is maintained using funds provided by the ANID-BASAL Center FB210003, CATA. This work is based [in part] on observations made with the NASA/ESA/CSA James Webb Space Telescope. The data were obtained from the Mikulski Archive for Space Telescopes at the Space Telescope Science Institute, which is operated by the Association of Universities for Research in Astronomy, Inc., under NASA contract NAS 5-03127 for JWST. These observations are associated with programs $\#$1201 and $\#$3231. JSJ gratefully acknowledges support from the FONDECYT grant 1240738 and from the ANID BASAL project FB210003.

\bibliography{ms}

\section*{Appendix}

\renewcommand{\thefigure}{A\arabic{figure}}
\renewcommand{\thetable}{A\arabic{table}}
\renewcommand{\theequation}{A\arabic{equation}}
\renewcommand{\thepage}{A\arabic{page}}
\setcounter{figure}{0}
\setcounter{table}{0}
\setcounter{equation}{0}
\setcounter{page}{1}

\begin{table}[h]
	\centering
	\caption{Priors of the free parameters used in the atmospheric retrievals. \label{tab:retrieval_priors}}
	\begin{tabular}{l c}
		\hline
		Parameter & Prior \\
		\hline
		$\mathrm{T_{int}} \text{ [K]}$ & $\mathcal{U}(0,2000)$\\
		$\mathrm{T_{eq}} \text{ [K]}$ & $\mathcal{U}(0,5000)$\\
		$[\gamma]$ & $\mathcal{U}(-2,2)$\\
		$[\kappa_{\mathrm{IR}}]$ & $\mathcal{U}(-4,0)$\\
        $\text{T}_0 \text{ [K]}$ & $\mathcal{U}(500, 3000)$ \\
        $\delta_{\mathrm{P}_{1}}$ & $\mathcal{U}(0, 0.8)$ \\
        $\delta_{\mathrm{P}_{2}}$ & $\mathcal{U}(0, 1)$ \\
        $\delta_{\mathrm{P}_{3}}$ & $\mathcal{U}(0, 1)$ \\
        $[\gamma_1]$ & $\mathcal{U}(-4, 1)$ \\
        $[\gamma_2]$ & $\mathcal{U}(-4, 1)$ \\
        $[\gamma_3]$ & $\mathcal{U}(-4, 1)$ \\
        $\xi_1$ & $\mathcal{U}(3, 6)$ \\
        $\xi_2$ & $\mathcal{U}(1, 5)$ \\
        $\xi_3$ & $\mathcal{U}(1, 5)$ \\
        $f_{\mathrm{s, NIRISS}}$ & $\mathcal{N}(1,0.001)$\\
        $f_{\mathrm{s, NRS2}}$ & $\mathcal{N}(1,0.001)$\\
		$f_{\mathrm{sed}}$ & $\mathcal{U}(2,12)$\\
		$\sigma$ & $\mathcal{U}(1,5)$\\
		$[K_{zz}]$ & $\mathcal{U}(2,15)$\\
		$f_c$ & $\mathcal{U}(0,1)$\\
		$[Z/Z_\odot]$ & $\mathcal{U}(-2,2)$\\
		$C/O$ & $\mathcal{U}(0.05,2)$\\
		$[P_q]$ & $\mathcal{U}(-12,2)$\\
		$[\mathrm{H_2O}]$ & $\mathcal{U}(-12,-0.5)$\\
		$[\mathrm{CO}]$ & $\mathcal{U}(-12, 0)$\\
		$[\mathrm{CO_2}]$ & $\mathcal{U}(-12,-0.5)$\\
		$[\mathrm{SiO}]$ & $\mathcal{U}(-14,-0.5)$\\
		$[\mathrm{CH_4}]$ & $\mathcal{U}(-14,-1.5)$\\
		$[\mathrm{C_2H_2}]$ & $\mathcal{U}(-14,-1.5)$\\
		$[\mathrm{HCN}]$ & $\mathcal{U}(-14,-1.5)$\\
		$[\mathrm{PH_3}]$ & $\mathcal{U}(-14,-1.5)$\\
		$[\mathrm{H_2S}]$ & $\mathcal{U}(-14,-1.5)$\\
		$[\mathrm{TiO}]$ & $\mathcal{U}(-14,-1.5)$\\
		$[\mathrm{VO}]$ & $\mathcal{U}(-14,-1.5)$\\
        $[\mathrm{Fe(s)}]$ & $\mathcal{U}(-14,0)$\\
		$[\mathrm{MgSiO_3(s)}]$ & $\mathcal{U}(-14,0)$\\
		$[\mathrm{Mg_2SiO_4(s)}]$ & $\mathcal{U}(-14,0)$\\
		$[\mathrm{Al_2O_3(s)}]$ & $\mathcal{U}(-14,0)$\\
		$[\mathrm{SiO_2(s)}]$ & $\mathcal{U}(-14,0)$\\
		$[\mathrm{CaTiO_3(s)}]$ & $\mathcal{U}(-14,0)$\\
		$[\mathrm{TiO_2(s)}]$ & $\mathcal{U}(-14,0)$\\
		\hline
	\end{tabular}
\end{table}

\begin{figure}
\includegraphics[width=1\columnwidth]{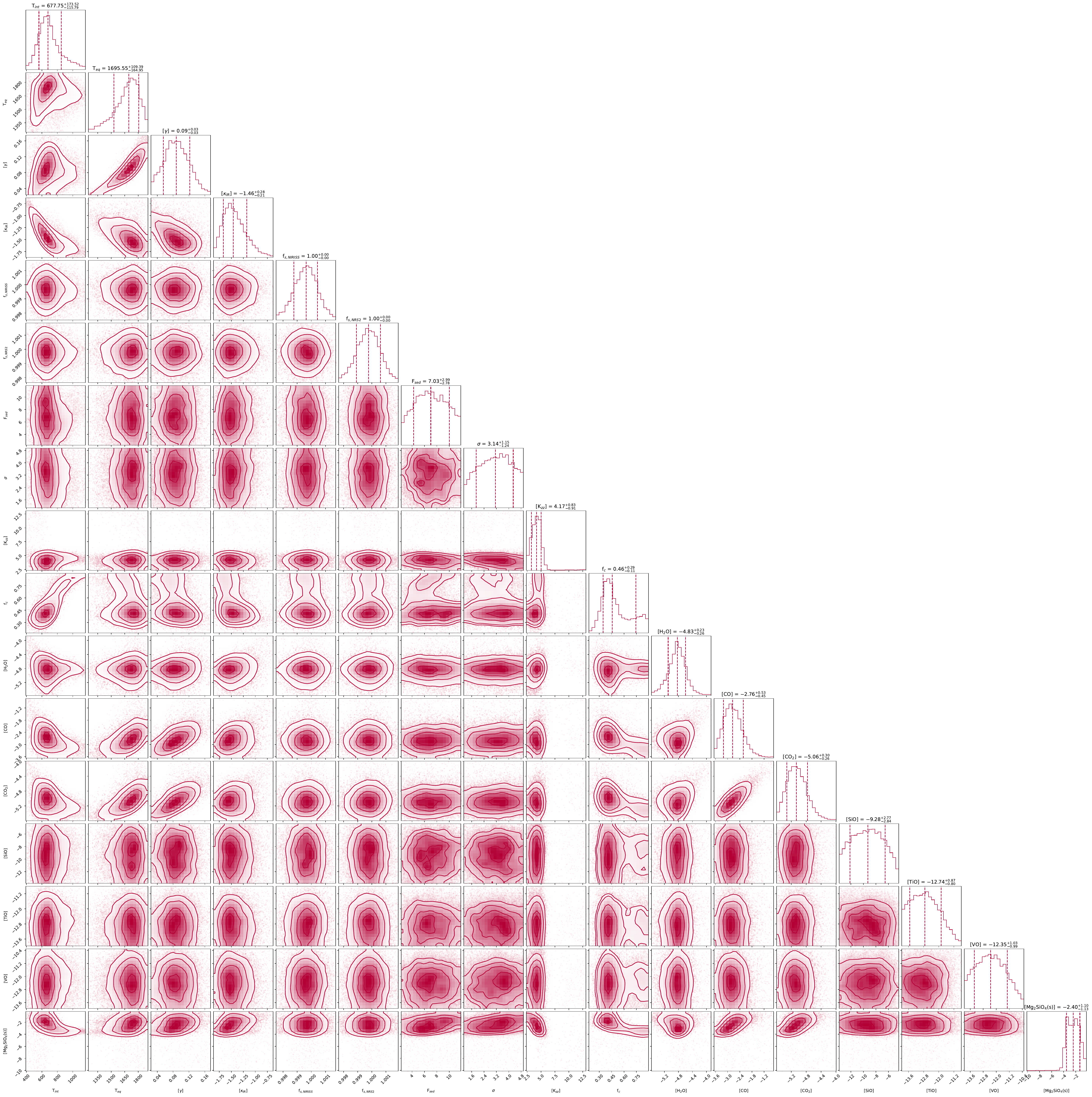}
\caption{Corner plot showing the posterior distributions from the free chemistry retrievals of the Eureka! spectrum using the Guillot P–T profile, including only the well-constrained molecular species. \label{fig:cor_eu}}
\end{figure}

\begin{figure}
\includegraphics[width=1\columnwidth]{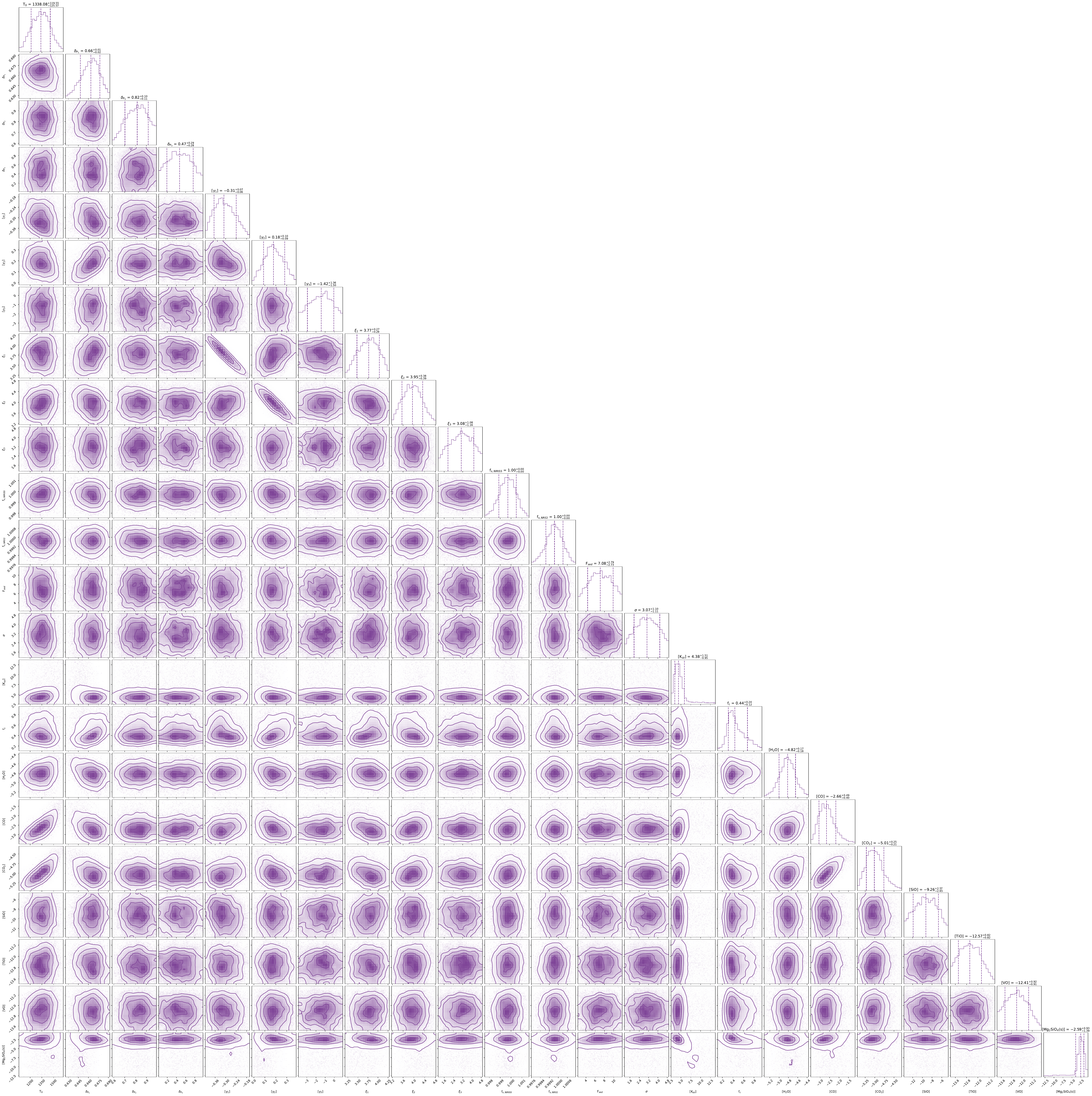}
\caption{Corner plot showing the posterior distributions from the free chemistry retrievals of the Eureka! spectrum using the neo-MS P–T profile, including only the well-constrained molecular species. \label{fig:m_cor_eu}}
\end{figure}

\begin{figure}
\includegraphics[width=1\columnwidth]{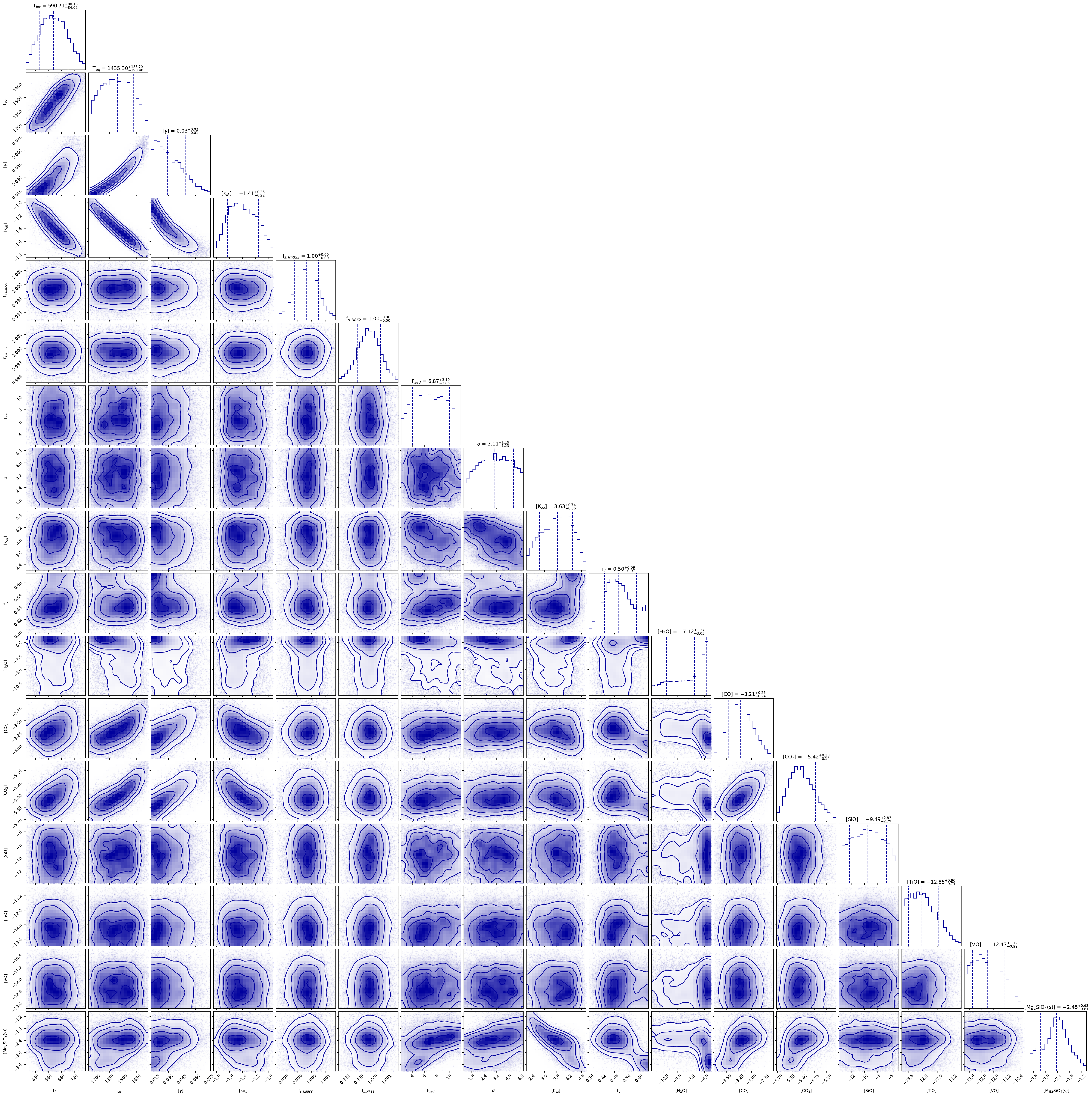}
\caption{Corner plot showing the posterior distributions from the free chemistry retrievals of the exoTEDRF spectrum using the Guillot P–T profile, including only the well-constrained molecular species. \label{fig:cor_et}}
\end{figure}

\begin{figure}
\centering
\includegraphics[width=\textwidth]{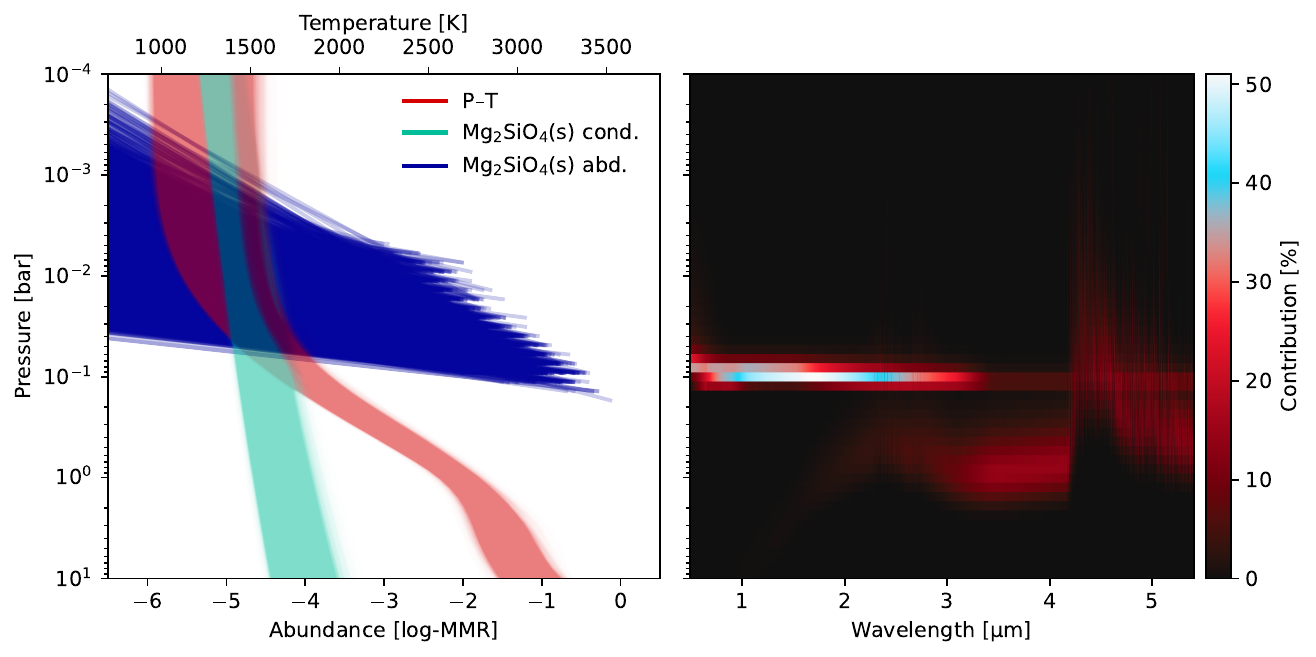}
\caption{Retrieved posterior P–T profiles from the free chemistry retrievals of the exoTEDRF spectrum, shown along with the corresponding posteriors for Mg$_2$SiO$_4$(s) condensation curves and abundance profiles, plus the median contribution functions.
\label{fig:contribution_et}}
\end{figure}

\begin{figure}
\centering
\includegraphics[width=\textwidth]{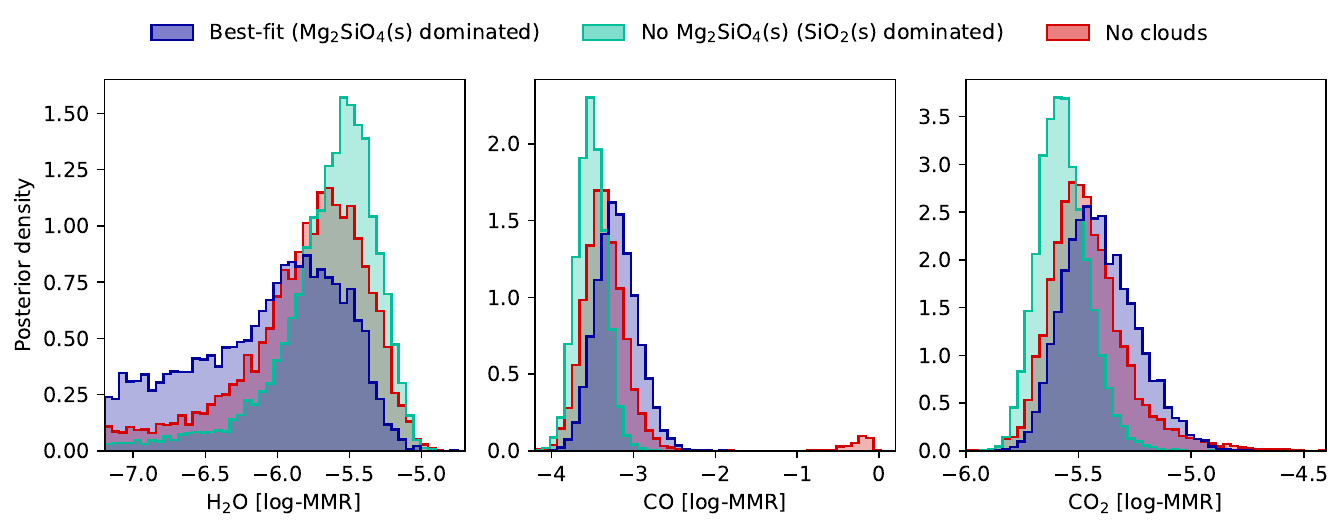}
\caption{Comparison of abundance posteriors for key molecular species from the free chemistry retrievals of the exoTEDRF spectrum.
\label{fig:comparison_exotedrf}}
\end{figure}

\end{document}